\documentclass[journal=jctcce,manuscript=article,layout=traditional]{achemso} 
\usepackage[T1]{fontenc}
\usepackage[version=3]{mhchem}
\usepackage[dvipsnames]{xcolor}
\usepackage{graphicx,subcaption}
\usepackage{amssymb}
\usepackage{overpic}
\usepackage{braket}
\usepackage{soul}
\usepackage{bm}
\usepackage{hyperref}
\usepackage{enumerate}
\usepackage{accents}
\hypersetup{pdfborderstyle={/S/U/W 0.5}}

\newcommand{\pp}[2]{\frac{\partial{#1}}{\partial{#2}}}
\newcommand{\dd}[2]{\frac{d {#1}}{d {#2}}}
\newcommand{\grad}[1]{\nabla_{\bm{#1}}}
\newcommand{\gradn}[1]{\nabla_{#1}}
\newcommand{\hp}[3]{\bm{H}^{{#1}}{:}\bm{P}^{{#2},{#3}}}
\newcommand{\ppip}[3]{\bm{P}^{{#2},{#3}}{:}\bm{\Pi}^{{#1}}{:}\bm{P}^{{#2},{#3}}}

\newcommand{\tbm}[1]{\tilde{\bm{#1}}}
\newcommand{\ubbm}[1]{\underaccent{\ \sim}{\bm{#1}}}
\newcommand{\bbm}[1]{\bar{\bm{#1}}}

\newcommand{\wt}[1]{{\bm{\Omega}\left({#1}\right)}}
\newcommand{\RR}[0]{{[\bm{R}]}}

\newcommand{\bhp}[3]{\bbm{H}^{{#1}}{:}\bbm{P}^{{#2},{#3}}}
\newcommand{\bppip}[3]{\bbm{P}^{{#2},{#3}}{:}\bbm{\Pi}^{{#1}}{:}\bbm{P}^{{#2},{#3}}}
\author{Tian Qiu}
\affiliation{Department of Chemistry, Princeton University, Princeton, NJ 08540, USA.}
\author{Joseph E. Subotnik}
\email{subotnik@princeton.edu}
\affiliation{Department of Chemistry, Princeton University, Princeton, NJ 08540, USA.}
\title[]{
Fast Methods For Multisite Charge Transfer Processes II:  Analytic Nuclear Gradients  and Nonadiabatic Dynamics For cCASSCF(1,M) and cCASSCF(2M-1,M) Wavefunctions}

\begin{document}
\maketitle

\begin{abstract}
We derive and implement   analytic nuclear gradients and derivative couplings for a constrained Complete Active Space Self-Consistent Field with a small  active space designed to model electron or hole transfer. Using a Lagrangian formalism, we are able to differentiate both the CASSCF energy and the constraint (which is required for globablly smooth surfaces), and the resulting efficient algorithm can be immediately applied to nonadiabatic dynamics simulations of charge transfer processes. Here, we run initial surface-hopping simulations of a  proton coupled electron transfer event for a phenoxyl-phenol system.
\end{abstract}

\section{Introduction}
Charge transfer (CT) processes play an important role in numerous chemical phenomena including redox reactions \cite{marcus:1956,Kurouski2015,Nazmutdinov2023}, photochemical processes \cite{Lee2023}, electrochemistry \cite{He2024:jacs}, and catalysis \cite{Zachman2022,Giulimondi2023}. Fundamental theoretical understanding of electronic coupling in these processes has been widely developed \cite{Newton:2003:electron_transfer,Lange:2008:CT,Dai:2022:excitation}. Their theoretical investigation is essential for advancing technologies in energy storage, molecular electronics, and designing of novel catalyst, requiring sophisticated theoretical frameworks to properly describe the electronic structure and dynamics involved.\cite{Nelson:2014:nonadiabatic_md,CrespoOtero:2018:nonadiabatic_dynamics,Fedorov:2019:nonadiabatic_dynamics} Despite their importance, accurate and efficient modeling of CT processes remains a challenge due to the complex interplay between electronic and nuclear motion, particularly near conical intersections where nonadiabatic effects become crucial\cite{Matsika:2011:conical_intersection}. On the one hand, single reference, mean-field theories such as Hartree Fock (HF) or density functional theory (DFT) often lack the accuracy to describe CT processes properly, as they fail to capture the multi-reference character of the wavefunction; Time-Dependent DFT (TD-DFT) has been widely applied to study CT processes,\cite{CarterFenk:2021:CT_TDDFT,Dai:2022:excitation} but the latter treats excited states and the ground state in an unbalanced fashion  (and cannot recover the correct topology of a crossing\cite{martinez:2006:ci_topology_wrong}). On the other hand, multi-reference methods such as Complete Active Space Self-Consistent Field (CASSCF) and Multi-Reference Configuration Interaction (MRCI) provide more accurate descriptions of electron correlation for CT processes.\cite{Szalay:2011:mrci,Lischka:2018:MR} However, while recent advances have expanded CASSCF capabilities through density matrix renormalization group (DMRG) theory and provided analytic derivative couplings for selected multi-state CAS implementations,  these approaches can be extremely computationally demanding, a fact which prohibits their practical application for most large scale dynamics simulations. This fundamental trade-off between accuracy and efficiency underscores the need for alternative methods that can be both reliable and computational efficient.

In a recent manuscript \cite{Qiu:2024:dsc,Qiu:2024:diis}, we have presented a dynamically weighted state-averaged  constrained CASSCF(1,M)/CASSCF(2M-1,M) approach, or  eDSC/hDSC for short, to accurately generate PES and wavefunctions for studying CT processes in radical systems. For a two-state problem, our current implementation requires only twice the cost of HF calculations, and for a $M$-state problem, the algorithm requires at most $M$ times the cost of HF calculations. Given its low cost  and balanced good accuracy, the development of analytic nuclear gradients and derivative couplings for this method has now become essential to enable its immediate application in nonadiabatic dynamics simulations. Gradients and derivative couplings are needed also for  efficient transition state searches as well as to  accurately capture electronic transitions at avoided crossings, which is critical for studying CT mechanisms with proper treatment of quantum effects.

With this background in mind, in this paper, we present the required mathematical formalism for computing such analytic nuclear gradients and  derivative couplings through a Lagrange multiplier approach.\cite{glover:2023:lagrange_multiplier,Athavale:2025:cis:gradient,Rappoport:2005} We have implemented the formalisms in QChem\cite{Epifanovsky2021} and verified these expressions by comparing them with finite-difference results. As a proof-of-concept, we have used our analytic nuclear gradients and derivative couplings to run surface hopping (SH) simulations of a proton-coupled electron transfer (PCET) process, demonstrating the practical utility of our approach. With  the current implementation, computing the analytic nuclear gradient and derivative coupling requires approximately the same computational cost as converging the electronic structure calculation within eDSC/hDSC for a system with 25 atoms and 232 basis functions (the phenoxyl-phenol system with 6-31g* basis set), which is only twice the cost of HF calculations. Thus, this work demonstrates the immediate feasibility of  nonadiabatic dynamics simulation for studying CT processes of large systems, a capability which should open the door to numerous research fields including photocatalysis, electrochemistry, transition metal catalysis, proton-coupled electron transfer, and the design of molecular electronics. 

An outline of this paper is as follows. In Sec. \ref{sec:grad_lag_general}, we briefly review (in general) the Lagrange multiplier approach for computing the analytic nuclear gradient. In Sec. \ref{sec:grad_lag_constrained}, we demonstrate how to apply such a theory to our constrained optimization problem. In Sec. \ref{sec:theory_ndsc}, we construct the relevant equations for  the $M$-state eDSC/hDSC problem. Specifically, in Secs. \ref{sec:ana_grad_ek}-\ref{sec:summary_dEkdR}, we derive the necessary terms to compute the analytic nuclear gradient of the relevant adiabatic energies and in Sec. \ref{sec:ana_dij}, we derive the necessary terms to for the derivative coupling. In Sec. \ref{sec:result}, we apply our formalism to SH simulations of a proton-coupled electron transfer process. Finally, in Sec. \ref{sec:conclusion} we summarize and point out new directions of this work.

\section{Background: Review of Analytic Nuclear Gradient through Lagrange Multiplier Approach}\label{sec:grad_lag_general}
Let $\{x_i\}$ be a set of independent variables and $\{R_i\}$ be a set of parameters. Let $g_i(\bm{x};\bm{R})$ be a set of constraints on $\bm{x}$ that depends parametrically on $\bm{R}$, where $\bm{x}$ is a shorthand for $(x_1,x_2,\cdots)^\top$ and $\bm{R}$ is shorthand for $(R_1,R_2,\cdots)^\top$. For clarity, one can think of $\bm{x}$ as some parameterization of the electronic wavefunction (e.g.,  the molecular orbital coefficients in the atomic orbital basis) while $\bm{R}$ represents the nuclear coordinates in a Hartree-Fock framework. A Lagrange multiplier approach provides a method to compute the total derivative of any target function w.r.t. $\bm{R}$ when working with fully determined systems \cite{Snyder2017,Paz2023}. To that end, suppose we have enough constraints $g_i(\bm{x};\bm{R})$ such that $\bm{x}$ can be determined locally by equations
\begin{align}
    g_i(\bm{x};\bm{R}) = 0.\label{eq:gi}
\end{align}
Then, for any arbitrary target function $f(\bm{x},\bm{R})$, one can build a Lagrangian
\begin{align}
    \mathcal{L}(\bm{x},\bm{R},\bm{\lambda}) = f(\bm{x},\bm{R})-\bm{\lambda}^\top\bm{g}(\bm{x},\bm{R}),
\end{align}
where $\bm{\lambda}=(\lambda_1,\lambda_2,\cdots)^\top$ is a collection of Lagrange multipliers and $\bm{g}(\bm{x},\bm{R}) = [g_1(\bm{x},\bm{R}),g_2(\bm{x},\bm{R}),\cdots]^\top$. According to a Lagrange multiplier approach, we can compute the total derivative of $f$ w.r.t. $\bm{R}$ on the constrained manifold (solution manifold of Eq. \ref{eq:gi}) by evaluating
\begin{align}
    \dd{f}{\bm{R}} = \grad{R}\mathcal{L}(\bm{x},\bm{R},\bm{\lambda}) = \grad{R}f(\bm{x},\bm{R})-\left(\grad{R}\bm{g}(\bm{x},\bm{R})\right)^\top\bm{\lambda}.\label{eq:dfdR}
\end{align}
Some words about  notation are now crucial. Throughout the paper, we will use $\grad{}$ to represent the gradient taken on the free manifold (i.e., all quantities $\bm{x}$, $\bm{\lambda}$, and $\bm{R}$ are independent of each other) and $\dd{}{\bm{R}}$ to represent the gradient taken on the constrained manifold (i.e., where $\bm{x}$ and $\bm{\lambda}$ depend on $\bm{R}$). With this convention, $\grad{R}$ in Eq. \ref{eq:dfdR} is taken with $\bm{x}$ and $\bm{\lambda}$ fixed, $\grad{R}\bm{g}(\bm{x},\bm{R})$ is a matrix indexed by
\begin{align}
    \left(\grad{R}\bm{g}(\bm{x},\bm{R})\right)_{ij} = \gradn{R_j}g_i(\bm{x},\bm{R}),
\end{align}
and $\bm{\lambda}$ is the solution to the equations $\grad{x}\mathcal{L}(\bm{x},\bm{R},\bm{\lambda}) = \bm{0}$, i.e.,
\begin{align}
    \bm{\lambda} = \left(\grad{x}\bm{g}(\bm{x},\bm{R})\right)^{-1\top}\grad{x}f(\bm{x},\bm{R}).\label{eq:lambda_solution}
\end{align}
Note that since $\bm{x}$ can be fully determined by equations $\bm{g}(\bm{x},\bm{R})=\bm{0}$, $\grad{x}\bm{g}(\bm{x},\bm{R})$ is a full-rank square matrix. With this approach, the total derivative of any arbitrary function $f(\bm{x},\bm{R})$ w.r.t. $\bm{R}$ can be computed (through Eqs. \ref{eq:dfdR} and \ref{eq:lambda_solution}) without computing $\dd{\bm{x}}{\bm{R}}$. Note that there is no mystery as far as deriving Eq. \ref{eq:dfdR}. The simplest means to derive this result is to write
\begin{align}
    \dd{f}{\bm{R}} = \grad{R}f+\left(\dd{\bm{x}}{\bm{R}}\right)^\top\grad{x}f.\label{eq:dfdR_expand}
\end{align}
And since $\dd{}{\bm{R}}$ is evaluated on the constrained manifold, one has
\begin{align}
    \bm{g}(\bm{x},\bm{R}) &= \bm{0}\\
    \Rightarrow \dd{\bm{g}(\bm{x},\bm{R})}{\bm{R}} &= \grad{R}\bm{g}(\bm{x},\bm{R}) + \grad{x}\bm{g}(\bm{x},\bm{R})\dd{\bm{x}}{\bm{R}} = \bm{0}\\
    \Rightarrow \dd{\bm{x}}{\bm{R}}&=-\left(\grad{x}\bm{g}(\bm{x},\bm{R})\right)^{-1}\grad{R}\bm{g}(\bm{x},\bm{R}).\label{eq:dxdR_expand}
\end{align}
 Eq. \ref{eq:dfdR} follows from 
 Eqs. \ref{eq:dfdR_expand} and \ref{eq:dxdR_expand} 
 provided that $\grad{x}\bm{g}(\bm{x},\bm{R})$ is invertible, i.e., the constraints determine the system.

\section{Theory: Analytic Nuclear Gradient for Constrained Optimization Problem}\label{sec:grad_lag_constrained}
The theory above is not general enough to handle most problems in quantum chemistry.  For electronic structure problems, it is often the case that, in order to find variables $\bm{x}$, we must first optimize a function $f(\bm{x},\bm{R})$  satisfying a constraint $g(\bm{x},\bm{R}) = 0$; thereafter, second, we need to differentiate a function $h(\bm{x},\bm{R})$. In other words,  rather than investigating a fully determined by a constraint, instead we must first solve a constrained optimization problem and then further differentiate. This case is clearly more complicated than the optimization treated above.

To study such a case, 
let the Lagrangian for the initial constrained optimization problem be
\begin{align}
    \mathcal{L}_1(\bm{x},\bm{R},\bm{\lambda}) = f(\bm{x},\bm{R})-\bm{\lambda}^\top\bm{g}(\bm{x},\bm{R}),
\end{align}
Note that the solution for $\bm{x}$ can be completely determined by the constraints that
\begin{align}
    \bm{g}(\bm{x},\bm{R}) &= \bm{0},\label{eq:l1_g}\\
    \grad{x}\mathcal{L}_1(\bm{x},\bm{R},\bm{\lambda}) &= \bm{0}.\label{eq:l1_dx}
\end{align}
Therefore, for a target function $h(\bm{x},\bm{R})$ that we wish to differentiate, we can build a second Lagrangian 
\begin{align}           
\mathcal{L}_2(\bm{x},\bm{R},\bm{\lambda},\bm{\mu},\bm{\nu}) = h(\bm{x},\bm{R}) - \bm{\mu}^\top \bm{g}(\bm{x},\bm{R}) - \bm{\nu}^\top \grad{x}\mathcal{L}_1(\bm{x},\bm{R},\bm{\lambda}),
\end{align}
such that $\bm{x}$ is now fully determined by the constraints that are contracted with the Lagrange multipliers $\bm{\mu}$ and $\bm{\nu}$. Following the approach discussed in Sec. \ref{sec:grad_lag_general}, the gradient of $h(\bm{x},\bm{R})$ can be evaluated by
\begin{align}
    \dd{h}{\bm{R}} = \grad{\bm{R}}\mathcal{L}_2 = \grad{R}h-(\grad{R}\bm{g})^\top \bm{\mu}-(\grad{R}\grad{x}\mathcal{L}_1)^\top\bm{\nu},\label{eq:dhdR}
\end{align}
where $(\grad{R}\grad{x}\mathcal{L}_1)$ is a matrix indexed by
\begin{align}
    (\grad{R}\grad{x}\mathcal{L}_1)_{ij} = (\gradn{R_j}\gradn{x_i}\mathcal{L}_1).
\end{align}
As always, here a partial derivative $\grad{}$ is taken by fixing the remaining four quantities  $\{\bm{x},\bm{R},\bm{\lambda},\bm{\mu},\bm{\nu}\}$. 

Now, all that remains is to determine Lagrange multipliers $\bm{\mu}$ and $\bm{\nu}$.  The crucial observation here is that, if we follow Sec. \ref{sec:grad_lag_general} above, there are actually two sets of equations for determining these Lagrange multipliers, namely:

\begin{align}
\grad{x}\mathcal{L}_2(\bm{x},\bm{R},\bm{\lambda},\bm{\mu},\bm{\nu}) &= \bm{0}.\label{eq:l2_dx}
\\    \grad{\lambda}\mathcal{L}_2(\bm{x},\bm{R},\bm{\lambda},\bm{\mu},\bm{\nu}) &= \bm{0}.\label{eq:l2_dlambda}
\end{align}

While Eq. \ref{eq:l2_dx} is naturally derived given that $\bm{x}$ is our independent variable in Eqs. \ref{eq:l1_g}-\ref{eq:l1_dx}, the key insight here is to notice that Eqs. \ref{eq:l1_g}-\ref{eq:l1_dx} determine not only $\bm{x}$ but also $\bm{\lambda}$;
to some extent, $\bm{\lambda}$ plays the same role as $\bm{x}$ in a constrained optimization problem which leads to Eq. \ref{eq:l2_dlambda}.
In any event,
combining Eqs. \ref{eq:l2_dx} and \ref{eq:l2_dlambda}, the equations for $\bm{\mu}$ and $\bm{\nu}$ are
\begin{align}
    \begin{pmatrix}
        \grad{xx}^2\mathcal{L}_1 &(\grad{x}\bm{g})^\top\\
        \grad{x}\bm{g} &\bm{0}
    \end{pmatrix}
    \begin{pmatrix}
        \bm{\nu}\\ \bm{\mu}
    \end{pmatrix}
    =
    \begin{pmatrix}
        \grad{x}h\\ \bm{0}
    \end{pmatrix}\label{eq:eqs_munu}.
\end{align}

Interestingly, Eq. \ref{eq:eqs_munu} is very similar to the Newton-KKT equations \cite{Nocedal2006}. Once $\bm{\mu}$ and $\bm{\nu}$ are solved from this equation, the gradient of the target function $h(\bm{x},\bm{R})$ can be evaluated by Eq. \ref{eq:dhdR} without computing $\dd{\bm{x}}{\bm{R}}$ explicitly.

\section{Brief Review of  eDSC/hDSC and Notation Used Below}\label{sec:theory_ndsc}
Let us now show how to apply the standard Lagrange multiplier approach  above to compute the analytic nuclear gradient of the adiabatic energies and  derivative couplings for constrained dynamically weighted state-averaged self-consistent field calculations with either $(i)$ one electron in $M$ orbitals (corresponding to ``electron transfer'') or $2M-1$ electrons in $2M$  orbitals (corresponding to ``hole transfer'').  
While one might refer to these methods as cDW-SA-CASSCF$(1,M)$ or cDW-SA-CASSCF$(2M,2M-1)$ calculations, henceforward we will use the even simpler acronyms eDSC and hDSC (for electron/hole \underline{d}ynamically weighted, \underline{s}tate averaged, \underline{C}ASSCF) \cite{Qiu:2024:dsc,Qiu:2024:diis}. Note that eDSC/hDSC use restricted orbitals, so that the spin up and spin down orbitals are equivalent spatially. Furthermore, to avoid the complexity of notation from indices of vector/matrix/tensor elements, here we introduce several shortcuts for notation used below.
\begin{itemize}
    \item Einstein summation convention is applied when the same indices corresponding to vector/matrix/tensor elements appear twice.
    \item A contraction notation (${:}$) is introduced such that for any matrix $\bm{A}$ and $\bm{B}$, we define $\bm{A}{:}\bm{B} = A_{ij}B_{ij} = {\rm Tr}(\bm{A}^\top\bm{B})$.
    \item The contraction notation is also applied between any matrix $\bm{A}$ and rank-4 tensor $\bm{\Pi}$ such that $\bm{\Pi}{:}\bm{A}$ is a matrix indexed by $(\bm{\Pi}{:}\bm{A})_{ij} = \Pi_{ijkl}A_{kl}$, and $\bm{A}{:}\bm{\Pi}$ is a matrix indexed by $(\bm{A}{:}\bm{\Pi})_{kl} = A_{ij}\Pi_{ijkl}$.
    \item The contraction between any two rank-4 tensors $\bm{\Pi}$ and $\bm{\Theta}$ is also a rank-4 tensor, which is indexed by $(\bm{\Pi}{:}\bm{\Theta})_{ijkl} = \Pi_{ijmn}\Theta_{mnkl}$.
    \item For given vector $\bm{V}$ indexed by $pq$ (with  elements $V_{pq}$ where $p <q$), it will be helpful to construct the corresponding antisymmetric matrix.  To that end, we define $\ubbm{V}$ as the matrix with elements:
\begin{align}
    (\ubbm{V})_{pq} = \begin{cases}
        V_{pq}, &p<q,\\
        0, &p=q,\\
        -V_{qp}, &p>q.
    \end{cases}\label{eq:V}
\end{align}
\end{itemize}
As far as matrices and basis sets for quantum chemistry are concerned, the one-electron Hamiltonian $\bm{H}^0$, the two-electron interaction tensors $\bm{\Pi}^J$ (Coulomb) and $\bm{\Pi}^K$ (exchange), and all density matrices $\bm{P}$s are expressed in an atomic orbital (AO) basis, $\{\ket{\chi_\rho}\}$. All quantities that are  explicitly contracted with these quantities are also expressed in an AO basis. There will be occasions below where matrices are expressed in a molecular orbital (MO) basis; 
for the most part, such matrices are written with a tilde, $\tbm{o}$.  
Finally, we will use a bar, $\bbm{o}$, to indicate a matrix expressed in an orthonormalized AO basis (and we will  of course define the basis transformation explicitly when introducing such matrices).
There is no distinction between upper and lower case variables. Lastly, as far as matrix/tensor indices are concerned, we use $\rho,\sigma,\zeta,\eta$ for indices in an AO basis, $p,q,r,s$ for indices in a MO basis, and $i,j,k,l,m,n$ for  general indices.

At this point, let us briefly review the eDSC/hDSC formalism. The readers may refer to  Refs. \citenum{Qiu:2024:dsc} and \citenum{Qiu:2024:diis} for more details. For an $M$ state problem, the Lagrangian is
\begin{align}
    \mathcal{L}_1 = \sum_{i=1}^{M} w_i E_i - \sum_{j=1}^{M-1}\lambda_j\bm{Q}^j{:}\bm{P}^{\rm active}\label{eq:lag1}
\end{align}
where $w_i$ is the dynamical weighting factor, $E_i$ is the energy of the $i{\rm th}$ electron configuration, $\lambda_j$ is the Lagrange multiplier for the $j{\rm th}$ constraint, $\bm{Q}^j$ is the kernel of the $j{\rm th}$ constraint ($\bm{Q}_j$ is a matrix that depends only on the nuclear geometry $\bm{R}$), and $\bm{P}^{\rm active}$ is the  projector onto the vector space spanned by all active orbitals (or equivalently, the density matrix from all active orbitals). For concreteness, here we will review each term in detail.

\begin{itemize}
\item The dynamical weighting factors are defined by
\begin{align}
    w_j &= \frac{a_j}{\sum_j a_j}\label{eq:w_first},\\
    a_j &= \frac{1-e^{-\Delta E_j/T}}{\Delta E_j/T},\\
    \Delta E_j &= E_j - E_1,\label{eq:w_last}
\end{align}
where $T$ is a fixed ``temperature'' parameter and configuration 1 is assumed to have the lowest energy among all electron configurations.

\item The energy for state $i$ can be written as
\begin{align}
    E_i = \frac{1}{2}\left(2\hp{0}{i}{t}+\ppip{J}{i}{t}-\ppip{K}{i}{\alpha}-\ppip{K}{i}{\beta}\right),\label{eq:ei}
\end{align}
where $\bm{H}^0$ is the one-electron Hamiltonian, $\bm{\Pi}^J$ is the Coulomb interaction tensor, 
\begin{equation}
    \Pi_{\rho\sigma\zeta\eta}^J = \int d\bm{r}^1d\bm{r}^2\chi_{\rho}(\bm{r}^1)\chi_{\sigma}(\bm{r}^1)\frac{1}{|\bm{r}^1-\bm{r}^2|}\chi_{\zeta}(\bm{r}^2)\chi_{\eta}(\bm{r}^2),
\end{equation}
$\bm{\Pi}^K$ is the exchange interaction tensor,
\begin{equation}
    \Pi_{\rho\sigma\zeta\eta}^K = \int d\bm{r}^1d\bm{r}^2\chi_{\rho}(\bm{r}^1)\chi_{\zeta}(\bm{r}^1)\frac{1}{|\bm{r}^1-\bm{r}^2|}\chi_{\sigma}(\bm{r}^2)\chi_{\eta}(\bm{r}^2),
\end{equation}

\item
$\bm{P}^{i,\alpha}$ is the spin-up density matrix of the $i{\rm th}$ electron configuration, $\bm{P}^{i,\beta}$ is the spin-down density matrix of the $i{\rm th}$ electron configuration, and $\bm{P}^{i,t}$ is the total density matrix summed over both spins, i.e., $\bm{P}^{i,t} = \bm{P}^{i,\alpha}+\bm{P}^{i,\beta}$.
\end{itemize}

The idea behinds eDSC/hDSC is to constrain the vector space spanned by all active orbitals projects equally onto each atomic fragment (charge center) so as to treat electron and hole transfer. To achieve this feat, we write
\begin{align}
    \bm{Q}^j = \bm{\Sigma}^j - \bm{\Sigma}^{j+1},
\end{align}
where $\bm{\Sigma}^j$ defines the projection onto the vector space spanned by the $j$th atomic fragment. In the atomic orbital (AO) basis, we find that it is convenient to express 
\begin{align}
    \bm{\Sigma}^j &= 
     \begin{pmatrix}
         \bm{S}^{11} &  \cdots & \bm{S}^{1n}\\
         \vdots &  \ddots   & \vdots \\
         \bm{S}^{n1} &   \cdots   & \bm{S}^{nn}
    \end{pmatrix}
    \begin{pmatrix}
        0  &\cdots &0\\
        \vdots  &\left(\bm{S}^{jj,{\rm frag}}\right)^{-1} &\vdots\\
        0  &\cdots &0
    \end{pmatrix}
     \begin{pmatrix}
         \bm{S}^{11} &  \cdots & \bm{S}^{1n}\\
         \vdots &  \ddots   & \vdots \\
         \bm{S}^{n1} &   \cdots   & \bm{S}^{nn}
    \end{pmatrix}
    \\
    & \equiv \bm{S}     
        \begin{pmatrix}
        0  &\cdots &0\\
        \vdots  &\left(\bm{S}^{jj,{\rm frag}}\right)^{-1} &\vdots\\
        0  &\cdots &0
        \end{pmatrix}\bm{S}
\end{align}
where $\bm{S}$ is the overlap matrix between AOs and $\bm{S}^{jj,{\rm frag}}$ is the diagonal block of $\bm{S}$ that belongs to the $j$th atomic fragment (charge center). Note that if the $j$th fragment includes all atoms, $\bm{\Sigma}^j = \bm{S}$ and ${\rm Tr}(\bm{\Sigma}^j\bm{P}^{\rm active}) = M$.

Now, when choosing the property  independent variable for optimization, note that the density matrices are suboptimal because of complications from the orthonormality constraints. Instead, a cleaner strategy is to parameterize the Lagrangian in Eq. \ref{eq:lag1} by the generator of the rotation matrix for the molecular orbitals (MO). Thus, we write
\begin{align}
    \bm{P}^{i,\alpha} &= \bm{C}\bm{N}^{i,\alpha}\bm{C}^\top,\\
    \bm{P}^{i,\beta} &= \bm{C}\bm{N}^{i,\beta}\bm{C}^\top,\\
    \bm{P}^{\rm active} &= \bm{C}\bm{N}^{\rm active}\bm{C}^\top,
\end{align}
and
\begin{align}
    \bm{C} = \bm{C}^{\rm old}e^{\bm{A}},\label{eq:c2a}
\end{align}
Here, the $\bm{N}$ matrices are occupation matrices, which are diagonal with unity diagonal elements if the orbital is occupied and zero otherwise. $\bm{C}$ is the MO coefficient and $\bm{A}$ is an anti-symmetric generator matrix. With this parameterization, the $\grad{x}$ operator in Eq. \ref{eq:dhdR} can be understood as $\grad{A}$.

 Let us now derive the working equations for the analytic nuclear gradient of adiabatic energies  (Secs. \ref{sec:ana_grad_ek}-\ref{sec:summary_dEkdR}) and the derivative couplings (Sec. \ref{sec:ana_dij}).
Obviously, the key obstacle is to extract the necessary lagrange multipliers, which we do now.

\section{Extracting Lagrange Multipliers (Using Eq. \ref{eq:eqs_munu}) }\label{sec:ana_grad_ek}
In the context of eDSC/hDSC calculations, if we seek the analytic nuclear gradient of the $k$th adiabatic energy ($E_k$), Eq. \ref{eq:eqs_munu} reads as follows
\begin{align}
    \begin{pmatrix}
        \grad{AA}^2E_{\rm tot}-\grad{AA}^2G_0 &(\grad{A}\bm{G})^\top\\
        \grad{A}\bm{G} &\bm{0}
    \end{pmatrix}
    \begin{pmatrix}
        \bm{\nu}\\ \bm{\mu}
    \end{pmatrix}
    =
    \begin{pmatrix}
        \grad{A}E_k\\ \bm{0}
    \end{pmatrix}\label{eq:munu_grad_adiabat}
\end{align}
where
\begin{align}
    E_{\rm tot} &= \sum_{i=1}^{M}w_iE_i,\\
    G_0 &= \sum_{j=1}^{M-1}\lambda_j\bm{Q}^j{:}\bm{P}^{\rm active},\\   
    G_{j>0}&=\bm{Q}^j{:}\bm{P}^{\rm active},\label{eq:G_j}\\
    \bm{G} &= (G_1,G_2,\cdots,G_{M-1})^\top.
\end{align}
Let us begin by  evaluating the terms on the L.H.S of Eq. \ref{eq:munu_grad_adiabat}.

\subsection[Energy Term]{Term $\grad{AA}^2E_{\rm tot}$}\label{sec:dEAA}
We show in Appendix \ref{sec:Appendix:wi} that, given the definition of $w_i$ in Eqs. \ref{eq:w_first}-\ref{eq:w_last}, one can write
\begin{align}
    \pp{^2E_{\rm tot}}{A_{rs}A_{pq}} = \pp{^2E_1}{A_{rs}A_{pq}} + \sum_{ij>1}^Mz_{ij}\pp{\Delta E_i}{A_{rs}}\pp{\Delta E_j}{A_{pq}}+\sum_{j>1}^Mb_j'\pp{^2\Delta E_j}{A_{rs}A_{pq}},\label{eq:dEdAA}
\end{align}
where
\begin{align}
    z_{i\neq j} &= -\frac{a_i'b_j'+a_j'b_i'}{\sum_ja_j},\\
    z_{ii} &=\frac{-e^{\Delta E_i/T}/T-a_i''\Delta E-2a_i'b_i'}{\sum_ja_j},\\
    a_i'' &=-\frac{e^{-\Delta E_i/T}/T+2a_i'}{\Delta E_i},\\
    b_{i>1}' &=\frac{e^{-\Delta E_i/T}-a_i'\Delta E}{\sum_j a_j},\\
    a_i' &= \frac{e^{-\Delta E_i/T}-a_i}{\Delta E_i},\\
    \Delta E &= E_{\rm tot}-E_1.
\end{align}
and $E_i$ is defined in Eq. \ref{eq:ei} above.
Again, we assume that $E_1$ is the lowest state energy. According to Eq. \ref{eq:dEdAA}, we can evaluate $\grad{AA}^2E_{\rm tot}$ once we compute $\grad{AA}^2E_i$ and $\grad{A}E_i$.  In order to calculate the latter two terms, recalling Eq. \ref{eq:ei}, let us use the notation
\begin{align}
    \epsilon_1(\bm{H},\bm{P}) &= \bm{H}{:}\bm{P},\label{eq:one_ele_ene}\\
    \epsilon_2(\bm{\Pi},\bm{P})&= \bm{P}{:}\bm{\Pi}{:}\bm{P},
\end{align}
such that
\begin{align}
    E_i = \epsilon_1(\bm{H}^0,\bm{P}^{i,t})+\frac{1}{2}\left(\epsilon_2(\bm{\Pi}^J,\bm{P}^{i,t})-\epsilon_2(\bm{\Pi}^K,\bm{P}^{i,\alpha})-\epsilon_2(\bm{\Pi}^K,\bm{P}^{i,\beta})\right).
\end{align}
It is then straightforward to differentiate and we   find (for $q>p$, $s>r$)
\begin{align}
    \left.\pp{\epsilon_1(\bm{H},\bm{P})}{A_{pq}}\right|_{\bm{A}=\bm{0}} =~&2\left([\tbm{H},\bm{N}]\right)_{pq},\\
    \left.\pp{^2\epsilon_1(\bm{H},\bm{P})}{A_{rs}A_{pq}}\right|_{\bm{A}=\bm{0}} =~&2\left(\tilde{W}_{ps}\delta_{qr}+\delta_{ps}\tilde{W}_{qr}-\tilde{W}_{pr}\delta_{qs}-\delta_{pr}\tilde{W}_{qs}\right)\nonumber\\
    &-2\left(\tilde{H}_{ps}N_{qr}+N_{ps}\tilde{H}_{qr}-\tilde{H}_{pr}N_{qs}-N_{pr}\tilde{H}_{qs}\right),
\end{align}
where $[\bm{A},\bm{B}]=\bm{AB}-\bm{BA}$ is the commutator between two matrices $\bm{A}$ and $\bm{B}$, and
\begin{align}
    \tbm{H} &= \bm{C}^\top\bm{HC},\\
    \label{eqn:W}
    \tbm{W} &= \frac{1}{2}(\tbm{H}\bm{N}+\bm{N}\tbm{H}).
\end{align}
In order to further simplify the notation that will be used later, one may also define a rank-4 tensor $\bm{\Theta}^1$
\begin{align}
    \left(\bm{\Theta}^1(\tbm{H},\bm{N})\right)_{pqrs}=
    \begin{cases}
        2\left(\tilde{W}_{ps}\delta_{qr}+\delta_{ps}\tilde{W}_{qr}-\tilde{W}_{pr}\delta_{qs}-\delta_{pr}\tilde{W}_{qs}\right)
    \label{joe_eqn}\\
\; \; \; \; - 2\left(\tilde{H}_{ps}N_{qr}+N_{ps}\tilde{H}_{qr}-\tilde{H}_{pr}N_{qs}-N_{pr}\tilde{H}_{qs}\right)   &\mbox{if} \; q>p, s>r\\
    0,&{\rm otherwise}
    \end{cases}
\end{align}
(using Eq. \ref{eqn:W} for $\tbm{W}$) such that
\begin{align}
\left.\pp{^2\epsilon_1(\bm{H},\bm{P})}{A_{rs}A_{pq}}\right|_{\bm{A}=\bm{0}} = \left(\bm{\Theta}^1(\tbm{H},\bm{N})\right)_{pqrs}.\label{eq:de1daa}
\end{align}
Similarly, for $\epsilon_2(\bm{\Pi},\bm{P})$, one finds
\begin{align}
    \left.\pp{\epsilon_2(\bm{\Pi},\bm{P})}{A_{pq}}\right|_{\bm{A}=\bm{0}} =~&4\left([\tbm{B},\bm{N}]\right)_{pq},\\
    \left.\pp{^2\epsilon_2(\bm{\Pi},\bm{P})}{A_{rs}A_{pq}}\right|_{\bm{A}=\bm{0}} =~&2\left(\bm{\Theta}^1(\tbm{B},\bm{N})\right)_{pqrs}\nonumber\\
    &+2(\tilde{\Pi}_{pqrs}+\tilde{\Pi}_{pqsr}+\tilde{\Pi}_{qprs}+\tilde{\Pi}_{qpsr})(N_q-N_p)(N_s-N_r),\label{eq:de2dAA_v0}
\end{align}
where no summation is applied on $p,q,r,s$ and we have defined:
\begin{align}
    \bm{B} &= \bm{\Pi}{:}\bm{P},\label{eq:B_first}\\
    \tbm{B}&= \bm{C}^\top \bm{BC},\label{eq:B_last}\\
    \tilde{\Pi}_{pqrs}&=C_{p\rho}^{\top}C_{q\sigma}^\top\Pi_{\rho\sigma\zeta\eta}C_{\zeta r}C_{\eta s}.
\end{align}
Lastly, one can define another rank-4 tensor $\bm{\Theta}^2$
\begin{align}
    \left(\bm{\Theta}^2(\tbm{\Pi},\bm{N})\right)_{pqrs} = 
    \begin{cases}
        (\tilde{\Pi}_{pqrs}+\tilde{\Pi}_{pqsr}+\tilde{\Pi}_{qprs}+\tilde{\Pi}_{qpsr})(N_q-N_p)(N_s-N_r), & q>p,s>r\\
        0,&{\rm otherwise},
    \end{cases}
\end{align}
which simplifies Eq. \ref{eq:de2dAA_v0} to
\begin{align}
    \left.\pp{^2\epsilon_2(\bm{\Pi},\bm{P})}{A_{rs}A_{pq}}\right|_{\bm{A}=\bm{0}} =2\left(\bm{\Theta}^1(\tbm{B},\bm{N})+\bm{\Theta}^2(\tbm{\Pi},\bm{N})\right)_{pqrs}.\label{eq:de2daa}
\end{align}

At this point, we can evaluate $\grad{A}E_i$ and  $\grad{AA}^2E_i$ for each adiabatic state. To make our final state-averaged expression most easy to express, let us further define several analogues of Eq. \ref{eq:B_first}-\ref{eq:B_last} above for each state $i$ (where we contract the two electron tensors and each state's density matrix, either in the AO or MO basis):
\begin{align}    
    \bm{B}^{J,i,t} &\equiv \bm{\Pi}^J{:}\bm{P}^{i,t},\\
    \bm{B}^{K,i,\alpha} &\equiv  \bm{\Pi}^K{:}\bm{P}^{i,\alpha},\\
    \bm{B}^{K,i,\beta} &\equiv  \bm{\Pi}^K{:}\bm{P}^{i,\beta}.
\end{align}
and
\begin{align}
    \tbm{B}^{J,i,t}&\equiv  \bm{C}^\top \bm{B}^{J,i,t}\bm{C},\\
    \tbm{B}^{K,i,\alpha}&\equiv \bm{C}^\top \bm{B}^{K,i,\alpha}\bm{C},\\
    \tbm{B}^{K,i,\beta}&\equiv  \bm{C}^\top \bm{B}^{K,i,\beta}\bm{C}.
\end{align}
The spin-up (spin-down) Fock matrices for the $i$th electron configuration in the MO basis are naturally defined as follows:
\begin{align}
    \tbm{F}^{i,\alpha} &=\tbm{H}^0+\tbm{B}^{J,i,t}-\tbm{B}^{K,i,\alpha},\\
    \tbm{F}^{i,\beta} &=\tbm{H}^0+\tbm{B}^{J,i,t}-\tbm{B}^{K,i,\beta}.
\end{align}
\subsubsection[dEidApq]{Final Form for $\pp{E_i}{A_{pq}}$}
With the notation and derivation above, the first derivative of the energy for the $i$th electronic configuration is
\begin{align}
    \pp{E_i}{A_{pq}} = 2\left([\tbm{F}^{i,\alpha},\bm{N}^{i\alpha}] + [\tbm{F}^{i,\beta},\bm{N}^{i\beta}]\right)_{pq}.\label{eq:dEiA}
\end{align}

\subsubsection[dEidApqrs]{Final Form for $\pp{^2E_i}{A_{rs}A_{pq}}$}
Finally, given Eqs. \ref{eq:de1daa} and \ref{eq:de2daa}, we can write
\begin{align}
    \grad{AA}^2E_i = &\bm{\Theta}^1(\tbm{F}^{i,\alpha},\bm{N}^{i,\alpha})+\bm{\Theta}^1(\tbm{F}^{i,\beta},\bm{N}^{i,\beta})\nonumber\\
    +&\bm{\Theta}^2(\tbm{\Pi}^J,\bm{N}^{i,t})-\bm{\Theta}^2(\tbm{\Pi}^K,\bm{N}^{i,\alpha})-\bm{\Theta}^2(\tbm{\Pi}^K,\bm{N}^{i,\beta}).\label{eq:dEiAA}
\end{align}

\subsubsection{A Scheme for Iterative Inversion of Eq. \ref{eq:munu_grad_adiabat}}
In practice, one cannot afford to invert the L.H.S. of Eq. \ref{eq:munu_grad_adiabat}. Rather, one must compute its contraction with a vector and solve iteratively. To that end, let us now derive an expression for the  contraction of Eq. \ref{eq:dEiAA} with an arbitrary vector $\bm{V}$ (with antisymmetric analogue $\ubbm{V}$ as in Eq. \ref{eq:V}). 

To begin with, note that, for an arbitary vector/matrix  $\bm{V}$/$\ubbm{V}$, one can easily verify that:
\begin{align}
    \bm{\Theta}^1(\tbm{H},\bm{N})_{pqrs}V_{rs} = 2\left(\left(\tbm{H}\ubbm{V}\bm{N}-(\tbm{H}\ubbm{V}\bm{N})^\top\right)-\left(\tbm{W}\ubbm{V}-(\tbm{W}\ubbm{V})^\top\right)\right)_{pq}.\label{eq:theta1_v}
\end{align}
Here, $\tbm{H}$ and $\bm{N}$ are obviously symmetric matrices, and
\begin{align}
    \tbm{W} = \frac{1}{2}(\tbm{H}\bm{N}+\bm{N}\tbm{H})\label{eq:theta1_w}
\end{align}
is also a symmetric matrix. One can further verify that
\begin{align}
    \bm{\Theta}^2(\tbm{\Pi},\bm{N})_{pqrs}V_{rs} = \left(\bm{N}(\tbm{Y}+\tbm{Y}^\top)-(\tbm{Y}+\tbm{Y}^\top)\bm{N}\right)_{pq},\label{eq:theta2_v}
\end{align}
where
\begin{align}
    \bm{Y} &=  \bm{\Pi}{:}\left(\bm{C}(\bm{N}\ubbm{V}-\ubbm{V}\bm{N})\bm{C}^\top\right),\\
    \tilde{\bm{Y}} &= \bm{C}^\top\bm{YC}.\label{eq:theta2_Y}
\end{align}
Given Eqs. \ref{eq:theta1_v}-\ref{eq:theta2_Y}, the path to computing $\left( \partial^2 E_i/ \partial A_{rs} \partial A_{pq}\right)  V_{rs}$ is clear. Namely, one calculates the following quantities in sequential order:
\begin{align}
    \tbm{D}'^{i,\alpha} &= \tbm{F}^{i,\alpha}\ubbm{V}\bm{N}^{i,\alpha},\label{eq:Dbar_start}\\
    \tbm{D}''^{i,\alpha} &= \frac{1}{2}(\tbm{F}^{i,\alpha}\bm{N}^{i,\alpha}+\bm{N}^{i,\alpha}\tbm{F}^{i,\alpha})\ubbm{V},\\
    \tbm{D}^i & = \tbm{D}'^{i,\alpha}+\tbm{D}'^{i,\beta}-\tbm{D}''^{i,\alpha}-\tbm{D}''^{i,\beta},\label{eq:Dbar}\\
    \bm{Y}^{J,i,t} &= \bm{\Pi}^J{:}(\bm{C}(\bm{N}^{i,t}\ubbm{V}-\ubbm{V}\bm{N}^{i,t})\bm{C}^\top),\label{eq:zbar_start}\\
    \bm{Y}^{K,i,\alpha} &= \bm{\Pi}^K{:}(\bm{C}(\bm{N}^{i,\alpha}\ubbm{V}-\ubbm{V}\bm{N}^{i,\alpha})\bm{C}^\top),\\
    \tbm{Z}^{J,i,t} &= \bm{N}^{i,t}\bm{C}^\top(\bm{Y}^{J,i,t}+\bm{Y}^{J,i,t\top})\bm{C},\\
    \tbm{Z}^{K,i,\alpha} &= \bm{N}^{i,\alpha}\bm{C}^\top(\bm{Y}^{K,i,\alpha}+\bm{Y}^{K,i,\alpha\top})\bm{C},\\
    \tbm{Z}^i &= \tbm{Z}^{J,i,t} - \tbm{Z}^{K,i,\alpha} - \tbm{Z}^{K,i,\beta},\label{eq:zbar}
\end{align}
so that, in the end,  we can write
\begin{align}
    \pp{^2E_i}{A_{rs}A_{pq}}V_{rs} = \left(2(\tbm{D}^i-\tbm{D}^{i\top})+(\tbm{Z}^i-\tbm{Z}^{i\top})\right)_{pq}.
\end{align}
and
\begin{align}
    \pp{E_i}{A_{pq}}V_{pq} = 2{\rm Tr}\left(\tbm{D}'^{i,\alpha}+\tbm{D}'^{i,\beta}\right) = 2{\rm Tr}\left(\tbm{D}^i\right).
\end{align}

\subsubsection[dEtdAv]{Final Form for a Contraction of $\pp{^2E_{\rm tot}}{A_{rs}A_{pq}}$ with $\bm{V}$}
If we define
\begin{align}
    b_1' = 1-\sum_{i>1}b_i',
\end{align}
recall Eq. \ref{eq:dEdAA}, we can finally write
\begin{align}
    \left(\grad{AA}^2E_{\rm tot}\right){:}\bm{V} =~&2\sum_{i,j>1}(\grad{A}E_i-\grad{A}E_1)z_{ij}{\rm Tr}\left(\tbm{D}^j-\tbm{D}^1\right)\nonumber\\
    &+2\sum_ib_i'(\tbm{D}^i-\tbm{D}^{i\top})+\sum_ib_i'(\tbm{Z}^i-\tbm{Z}^{i\top}).\label{eq:dEAAv}
\end{align}

\subsection[G Term]{Term $\grad{AA}^2G_0$}
The $G_0$ term can be treated in the same way as the one-electron energy term in Eq. \ref{eq:one_ele_ene}. If we define
\begin{align}
    \bm{Q}^0 = \sum_{j=1}^{M-1}\lambda_j\bm{Q}^j,
\end{align}
we can write
\begin{align}
    G_0 = \epsilon_1(\bm{Q}^0,\bm{P}^{{\rm active}}).
\end{align}
Now, by using Eqs. \ref{eq:de1daa} and \ref{eq:theta1_v}-\ref{eq:theta1_w}, we have
\begin{align}
    \left(\grad{AA}^2G_{0}\right){:}\bm{V} = 2\left(\tbm{Q}^0\ubbm{V}\bm{N}^{\rm active}-(\tbm{Q}^0\ubbm{V}\bm{N}^{\rm active})^\top\right)-2\left(\tbm{W}^Q\ubbm{V}-(\tbm{W}^Q\ubbm{V})^\top\right),\label{eq:dGAAv}
\end{align}
where
\begin{align}
    \tbm{Q}^0 &= \bm{C}^\top\bm{Q}^0\bm{C},\\
    \tbm{W}^Q &= \frac{1}{2}(\tbm{Q}^0\bm{N}^{\rm active}+\bm{N}^{\rm active}\tbm{Q}^0).
\end{align}

\subsection[Grad G]{Term $\grad{A}\bm{G}$}
In Eq. \ref{eq:munu_grad_adiabat},  we must make a distinction between the term $\grad{A}\bm{G}$ and the termn $(\grad{A}\bm{G})^\top$, as the two quantities contract with vectors in different ways. The term $(\grad{A}\bm{G})^\top$ will be discussed in the next section. For a vector $\bm{V}$ that has the same dimension as in Eqs. \ref{eq:dEAAv} and \ref{eq:dGAAv}, we can write
\begin{align}
    (\grad{A}\bm{G}){:}\bm{V} = \begin{pmatrix}
        (\grad{A}G_1){:}\bm{V}\\
        (\grad{A}G_2){:}\bm{V}\\
        \vdots\\
        (\grad{A}G_{M-1}){:}\bm{V}\\
    \end{pmatrix},\label{eq:dGAv_def}
\end{align}
where $G_j$ is defined in Eq. \ref{eq:G_j} and one has
\begin{align}
    \grad{A}G_j &= 2[\tbm{Q}^j,\bm{N}^{\rm active}],\\
    \tbm{Q}^j &= \bm{C}^\top\bm{Q}^j\bm{C}.
\end{align}
Each row on the R.H.S. of Eq. \ref{eq:dGAv_def} can be written as
\begin{align}
    (\grad{A}G_{j>0}){:}\bm{V} = 2{\rm Tr}\left(\tbm{Q}^j\ubbm{V}\bm{N}^{\rm active}\right).\label{eq:dGjA}
\end{align}

\subsection[Grad Gt]{Term $(\grad{A}\bm{G})^\top$}\label{sec:dGnu}
For a vector $\bm{\mu}$ that has the length of $M-1$, the contraction between $(\grad{A}\bm{G})^\top$ and $\bm{\mu}$ is given by
\begin{align}
    (\grad{A}\bm{G})^\top\bm{\mu} = \sum_{j=1}^{M-1}\grad{A}G_j\mu_j = 2\left[\sum_{j=1}^{M-1}\mu_j\tbm{Q}^j,\bm{N}^{\rm active}\right].\label{eq:dGdAmu}
\end{align}

This completes our analysis of \ref{eq:munu_grad_adiabat}; we have now a practical way to solve the necessary Lagrange multipliers $\bm{\mu}$ and $\bm{\nu}$.

\section{Explicit Gradient w.r.t. Nuclear Geometry (Using Eq. \ref{eq:dhdR}}\label{sec:grad_R_terms}
 
 At this point, we are in a position to evaluate  Eq. \ref{eq:dhdR} for the nuclear gradient of the energy state $k$. 
Let us write out Eq. \ref{eq:dhdR} as:
\begin{align}
    \dd{E_k}{\bm{R}} = \grad{R}E_k - (\grad{R}\bm{G})^\top\bm{\mu}-(\grad{R}\grad{A}E_{\rm tot})^\top\bm{\nu}+(\grad{R}\grad{A}G_0)^\top\bm{\nu}.\label{eq:dekdR}
\end{align}
Given $\bm{\nu}$ and $\bm{\mu}$, all that remains is to evaluate the four explicit  gradient terms, $\grad{R}E_k$, $\grad{R}\bm{G}$, $\grad{R}\grad{A}E_{\rm tot}$, and $\grad{R}\grad{A}G_0$.

Now, when evaluating the explicit gradient terms above, e.g. $\grad{R} E_k$,  care must be still be taken when work in an atomic orbital basis -- even though we have already derived the necessary Lagrange multipliers.  To understand the subtlety, note the target function is of the form $E_k(\bm{R},\bm{C}(\bm{R}))$ (where $\bm{C}$ is the matrix of MOs in an atomic orbital basis), but we have chosen above (in Sec. \ref{sec:theory_ndsc}) to substitute $E_k(\bm{R},\bm{C}(\bm{R}))$ with $E_k(\bm{R},\bm{A}(\bm{R}))$ using Eq. \ref{eq:c2a}.  One must ask: can we safely define (in the sense of Lagrange multiplier approach) as in Eq. \ref{eq:maybe}?
\begin{eqnarray}
\label{eq:maybe}
    \grad{\bm{R}} E_k = \left.\pp{E}{\bm{R}}\right|_{\bm{A}} 
\end{eqnarray}

Unfortunately, this definition cannot  be correct because $\bm{C}$ at different geometry $\bm{R}$ satisfies different orthonormality constraints:
\begin{align}
    \bm{C}^{1\top}\bm{S}(\bm{R}^1)\bm{C}^1 &= \bm{I},\label{eq:csc_1}\\
    \bm{C}^{2\top}\bm{S}(\bm{R}^2)\bm{C}^2 &= \bm{I}.\label{eq:csc_2}
\end{align}
where $\bm{R}^1$ and $\bm{R}^2$ are two nuclear geometries, $\bm{S}(\bm{R}^1)$ and $\bm{S}(\bm{R}^2)$ are the AO overlap matrices at these two geometries, and $\bm{C}^1$ and $\bm{C}^2$ are MO coefficients in AO basis at these two geometries. In other words, in practice, we should actually consider the target function to be $E_k(\bm{R},\bm{C}(\bm{R}),\bm{S}(\bm{R}))$. 

The distinction above indicates that, for a proper gradient, one must take into acccount the gradient of the nuclear overlap matrix, which leads to well-known Pulay forces in gradient theory. For our purposes,  we can derive the relevant corrections simply by 
orthonormalize the AO basis and write $\bm{C}$ in the {\bf orthonormal AO basis} as the independent variable. Specifically, one can define
\begin{align}
    \bbm{C}=\bm{TC},\label{eq:c_bar}
\end{align}
where $\bm{T} = \bm{S}^{1/2}$ is the square root of the AO overlap matrix $\bm{S}$, and treat $\bbm{C}$ as our independent variable (still parameterized by $\bm{A}$).
The advantage of this choice is that $\bbm{C}$ is orthonormalized in any geometry, i.e., $\bbm{C}^\top\bbm{C} = \bm{I}$, and there is no need to introduce an extra constraint term for the orthonormality of orbitals. Moreover, since $\grad{A}$ evaluates the response of MOs that does not depend on the basis in which MOs are expressed, changing $\bm{C}$ to $\bbm{C}$ does not influence any results derived above (one can verify that the AO basis or orthonormal AO basis are both contracted out in expressions for the $\grad{A}$ terms). Given this clarification of independent variables, it is also clear what quantities are considered ``explicitly geometry dependent'' in Eq. \ref{eq:dekdR}. One should express $E_k,\bm{G},E_{\rm tot}$, and $G_0$ in terms of $\bbm{C}$, and all quantities except $\bbm{C}$ (and the Lagrange multipliers $\bm{\lambda}$) are considered geometry dependent. We demonstrate this treatment using $E_k$ as an example below, the same approach applies to all terms in Eq. \ref{eq:dekdR}.

For $E_k$, we first write
\begin{align}
    E_k &= \frac{1}{2}\left(2\hp{0}{k}{t}+\ppip{J}{k}{t}-\ppip{K}{k}{\alpha}-\ppip{K}{k}{\beta}\right)\\
    &=\frac{1}{2}\left(2\bhp{0}{k}{t}+\bppip{J}{k}{t}-\bppip{K}{k}{\alpha}-\bppip{K}{k}{\beta}\right),
\end{align}
where
\begin{align}
    \bbm{P}^{k,\alpha} &= \bm{T}\bm{P}^{k,\alpha}\bm{T},\\
    \bbm{P}^{k,\beta} &= \bm{T}\bm{P}^{k,\beta}\bm{T},\\
    \bbm{P}^{k,t} &= \bbm{P}^{k,\alpha}+\bbm{P}^{k,\beta},\\
    \bbm{H}^0 &= \bm{X}\bm{H}^0\bm{X},\\
    \bbm{\Pi}^J &= (\bm{X}{\otimes}\bm{X}){:}\bm{\Pi}^J{:}(\bm{X}{\otimes}\bm{X}),\\
    \bbm{\Pi}^K &= (\bm{X}{\otimes}\bm{X}){:}\bm{\Pi}^K{:}(\bm{X}{\otimes}\bm{X}).
\end{align}
given that $\bm{X} = \bm{S}^{-1/2}$ and $\bm{X}{\otimes}\bm{X}$ is a rank-4 tensor indexed by
\begin{align}
    (\bm{X}{\otimes}\bm{X})_{\rho\sigma\zeta\eta} = X_{\rho\zeta}X_{\sigma\eta}.
\end{align}
Now, the explicit geometry gradient should be applied to $\bbm{H}^0$, $\bbm{\Pi}^J$, and $\bbm{\Pi}^K$. For instance,
\begin{align}
    \bbm{H}^{0\RR} = \bm{X}^\RR\bm{H}^0\bm{X}+\bm{X}\bm{H}^{0\RR}\bm{X}+\bm{X}\bm{H}^0\bm{X}^\RR,
\end{align}
where we have introduced a superscript $\RR$ indicating that the geometry gradient is applied to each element of the corresponding matrix. Note that the $\bm{H}^{0\RR}$ term itself involves two terms, that the change of Coulomb potential (and other possible one-electron potentials) and the change of the matrix element due to the change of a geometry dependent basis. With this setup, we will evaluate each term in Eq. \ref{eq:dekdR}.

\subsection[ddREk]{Term $\grad{R}E_k$}
One can verify that the explicit nuclear gradient for $E_k$ is
\begin{align}
    \grad{R}E_k =~&\frac{1}{2}(\bm{H}^{0\RR}+\bm{F}^{k,\alpha\RR}){:}\bm{P}^{k,\alpha}+\frac{1}{2}(\bm{H}^{0\RR}+\bm{F}^{k,\beta\RR}){:}\bm{P}^{k,\beta}\nonumber\\
    -&2(\bm{F}^{k,\alpha}\bm{P}^{k,\alpha}+\bm{F}^{k,\beta}\bm{P}^{k,\beta}){:}(\bm{XT}^\RR),\label{eq:dEkR}
\end{align}
where
\begin{align}
    \bm{F}^{k,\alpha\RR} &=\bm{H}^{0\RR}+\bm{\Pi}^{J\RR}{:}\bm{P}^{k,t}-\bm{\Pi}^{K\RR}{:}\bm{P}^{k,\alpha},\\
    \bm{F}^{k,\beta\RR} &=\bm{H}^{0\RR}+\bm{\Pi}^{J\RR}{:}\bm{P}^{k,t}-\bm{\Pi}^{K\RR}{:}\bm{P}^{k,\beta},
\end{align}
and $\bm{T^\RR}$ can be evaluated easily from $\bm{S}^\RR$, see Appendix \ref{sec:Appendix:TR}. 

\subsubsection[ddRGmu]{Term $(\grad{R}\bm{G})^\top\bm{\mu}$}
In a similar fashion, we can write
\begin{align}
    (\grad{R}\bm{G})^\top\bm{\mu} = \Big(\sum_{j=1}^{M-1}\mu_j\bm{Q}_j^\RR\Big){:}\bm{P}^{\rm active}-2\left(\Big(\sum_{j=1}^{M-1}\mu_j\bm{Q}_j\Big)\bm{P}^{\rm active}\right){:}(\bm{XT}^\RR).\label{eq:dGdRmu}
\end{align}

\subsubsection[ddRAG0]{Term $(\grad{R}\grad{A}G_0)^\top\bm{\nu}$}
The $(\grad{R}\grad{A}G_0)^\top\bm{\nu}$ term and the term $(\grad{R}\grad{A}E_{\rm tot})^\top\bm{\nu}$ in Eq. \ref{eq:dekdR} are slightly different from the former two terms but can be treated in the similar way if the contraction with the Lagrange multiplier $\bm{\nu}$ is performed first. By introducing a matrix $\ubbm{\nu}$ from $\bm{\nu}$ in the same way as expressed in Eq. \ref{eq:V}, we can write
\begin{align}
    (\grad{R}\grad{A}G_0)^\top\bm{\nu} = \bm{Q}^{0\RR}{:}\wt{\ubbm{\nu},\bm{N}^{\rm active}}-2\left(\bm{Q}^{0}\wt{\ubbm{\nu},\bm{N}^{\rm active}}\right){:}(\bm{XT}^\RR),\label{eq:dGdRAnu}
\end{align}
where
\begin{align}
    \wt{\bm{A},\bm{B}} = \bm{C}(\bm{AB}-\bm{BA})\bm{C}^\top
\end{align}
for any two matrices $\bm{A}$ and $\bm{B}$, and 
\begin{align}
    \bm{Q}^{0\RR} &= \sum_{j=1}^{M-1}\lambda_j\bm{Q}^{j\RR}.
\end{align}

\subsubsection[ddRAEt]{Term $(\grad{R}\grad{A}E_{\rm tot})^\top\bm{\nu}$}
Finally,
\begin{align}
    \left(\grad{R}\grad{A}E_{\rm tot}\right)^\top\bm{\nu}=~&2\sum_{i,j>1}(\grad{R}E_i-\grad{R}E_1)z_{ij}{\rm Tr}\left(\tbm{D}^j-\tbm{D}^1\right)\nonumber\\
    +&\sum_ib_i'(\grad{R}\grad{A}E_i)^\top\bm{\nu},\label{eq:dEtotdARnu}
\end{align}
where $\tbm{D}^j$ is defined in Eqs. \ref{eq:Dbar_start}-\ref{eq:Dbar} by replacing $\ubbm{V}$ with $\ubbm{\nu}$ and
\begin{align}
    (\grad{R}\grad{A}E_i)^\top\bm{\nu} =~&\bm{F}^{i,\alpha\RR}{:}\wt{\ubbm{\nu},\bm{N}^{i,\alpha}}+\bm{F}^{i,\beta\RR}{:}\wt{\ubbm{\nu},\bm{N}^{i,\beta}}\nonumber\\
    -~&2\left(\bm{F}^{i,\alpha}\wt{\ubbm{\nu},\bm{N}^{i,\alpha}}\right){:}(\bm{XT}^\RR)-2\left(\bm{F}^{i,\beta}\wt{\ubbm{\nu},\bm{N}^{i,\beta}}\right){:}(\bm{XT}^\RR)\nonumber\\
    +~&{\rm Tr}\left(\bm{C}\tbm{Z}^i\bm{C}^\top\bm{SXT}^{\RR}\right),\label{eq:dEidARnu}
\end{align}
where $\tbm{Z}^i$ is defined in Eqs. \ref{eq:zbar_start}-\ref{eq:zbar} by replacing $\ubbm{V}$ with $\ubbm{\nu}$.

\section{Summary of Working Equations for the Nuclear Gradient}\label{sec:summary_dEkdR}
Here we summarize the working equations for the analytic nuclear gradient of the adiabatic energy $E_k$. The expression is
\begin{align}
    \dd{E_k}{\bm{R}} = \grad{R}E_k - (\grad{R}\bm{G})^\top\bm{\mu}-(\grad{R}\grad{A}E_{\rm tot})^\top\bm{\nu}+(\grad{R}\grad{A}G_0)^\top\bm{\nu},\label{eq:dEdR_final}
\end{align}
where $\bm{\mu}$ and $\bm{\nu}$ are solved by Eq. \ref{eq:munu_grad_adiabat}
through iterative methods, given that $\grad{A}E_k$ is evaluated by Eq. \ref{eq:dEiA}; for any vector $\bm{V}$ of the same size as $\bm{\nu}$ (note that $\bm{V}$ is usually indexed by two indices $V_{pq},p<q$), $\left(\grad{AA}^2E_{\rm tot}\right){:}\bm{V}$ is evaluated by Eq. \ref{eq:dEAAv}, $\left(\grad{AA}^2G_{0}\right){:}\bm{V}$ is evaluated by Eq. \ref{eq:dGAAv}, $(\grad{A}G_{j>0}){:}\bm{V}$ is evaluated by Eq. \ref{eq:dGjA}; and for any vector $\bm{\mu}$ of length $M-1$, $(\grad{A}\bm{G})^\top\bm{\mu}$ is evaluated by Eq. \ref{eq:dGdAmu}.

After solving for $\bm{\mu}$ and $\bm{\nu}$, one can evaluate Eq. \ref{eq:dEdR_final} term by term, where $\grad{R}E_k$ is evaluated by Eq. \ref{eq:dEkR}, $(\grad{R}\bm{G})^\top\bm{\mu}$ is evaluated by Eq. \ref{eq:dGdRmu}, $(\grad{R}\grad{A}G_0)^\top\bm{\nu}$ is evaluated by Eq. \ref{eq:dGdRAnu}, and finally, $\left(\grad{R}\grad{A}E_{\rm tot}\right)^\top\bm{\nu}$ is evaluated by Eqs. \ref{eq:dEtotdARnu} and \ref{eq:dEidARnu}.

\section{Derivative Coupling}\label{sec:ana_dij}
Having derive the relevant equations for the energy gradient, let us turn our attention to the the derivative coupling between two adiabatic states.  A result can be found  with only minor revisions on the procedure proposed in the previous section. Let us define
\begin{align}
    \bm{d}^{kl} = \left<\Psi_k\right|\dd{}{\bm{R}}\left|\Psi_l\right>\label{eq:d_kl_v1}
\end{align}
as the derivative coupling between adiabatic states $\left|\Psi_k\right>$ and $\left|\Psi_l\right>$, where both $\left|\Psi_k\right>$ and $\left|\Psi_l\right>$ can be expressed as a single Slater determinant of all the occupied orbitals. Since different adiabatic states differ only by their active orbitals, Eq. \ref{eq:d_kl_v1} is equivalent to
\begin{align}
    \bm{d}^{kl} = \left<\phi_{a_k}\right|\dd{}{\bm{R}}\left|\phi_{a_l}\right>\label{eq:d_kl_v2}
\end{align}
where $\{\left|\phi\right>\}$ are one-electron orbitals and $a_l$ is the index of the $l{\rm th}$ active orbital. 

Now, in an AO basis $\{\left|\chi_{\rho}\right>\}$, any one-electron orbital can be expressed as
\begin{align}
    \left|\phi_i\right> = \sum_{\rho} C_{\rho i}\left|\chi_{\rho}\right>,
\end{align}
where $C_{\rho i}$ 
defines a matrix  of molecular orbitals, $\bm{C}$. Eq. \ref{eq:d_kl_v2} can then be expanded as
\begin{align}
    \bm{d}^{kl} &= \sum_{\rho\sigma}C_{\rho a_k}\braket{\chi_\rho|\dd{}{\bm{R}}\chi_\sigma}C_{\sigma a_l} + \sum_{\rho\sigma}C_{\rho a_k}\braket{\chi_\rho|\chi_\sigma}\dd{}{\bm{R}}C_{\sigma a_l}\\
    &=\sum_{\rho\sigma}C_{\rho a_k}\braket{\chi_\rho|\grad{R}\chi_\sigma}C_{\sigma a_l}+\sum_{\rho\sigma}C_{\rho a_k}S_{\rho\sigma}\dd{}{\bm{R}}C_{\sigma a_l}.\label{eq:d_kl_v3}
\end{align}
The first term on the R.H.S. of Eq. \ref{eq:d_kl_v3} is straightforward to compute; our goal is to compute the second term, i.e., $\sum_{\rho\sigma}C_{\rho a_k}S_{\rho\sigma}\dd{}{\bm{R}}C_{\sigma a_l}$.

Next, consider an analogous target function
defined in a basis of orthonormal atomic orbitals,
\begin{align}
    D^{kl} = \left(\bbm{C}^{0\top}\bbm{C}\right)_{a_ka_l},
\end{align}
where both $\bbm{C}^0$ and $\bbm{C}$ are defined in Eq. \ref{eq:c_bar} -- except that henceforward, $\bbm{C}^0$ is considered a constant matrix when taking any derivatives. One can easily show that
\begin{align}
    \dd{D^{kl}}{\bm{R}} &= \left(\bbm{C}^{0\top}\dd{\bm{T}}{\bm{R}}\bm{C}\right)_{a_ka_l} + \left(\bbm{C}^{0\top}\bm{T}\dd{\bm{C}}{\bm{R}}\right)_{a_ka_l}\\
    &=\left(\bm{C}^\top\bm{T}\bm{T}^\RR\bm{C}\right)_{a_ka_l} + \left(\bm{C}^{\top}\bm{S}\dd{\bm{C}}{\bm{R}}\right)_{a_ka_l}\label{eq:dDdR}
\end{align}
At this point, we notice that the first term of Eq. \ref{eq:dDdR} is again trivial to compute, and the second term is identical to $\sum_{\rho\sigma}C_{\rho a_k}S_{\rho\sigma}\dd{}{\bm{R}}C_{\sigma a_l}$ found above. Therefore, we may conclude that:
\begin{align}
    \bm{d}^{kl} = \dd{D^{kl}}{\bm{R}} - \left(\bm{C}^\top\bm{T}\bm{T}^\RR\bm{C}\right)_{a_ka_l} + \sum_{\rho\sigma}C_{\rho a_k}\braket{\chi_\rho|\grad{R}\chi_\sigma}C_{\sigma a_l}.\label{eq:d_kl_vfinal}
\end{align}

Eq. \ref{eq:d_kl_vfinal} offers us an approach to compute the derivative coupling using the formalism derived above in Sec. \ref{sec:summary_dEkdR}.  After all, the Lagrangian approach is general for computing the analytic nuclear gradient of any target function so that, in principle,  computing $\dd{D^{kl}}{\bm{R}}$ should be no more difficult than computing $\dd{E_k}{\bm{R}}$. 
In fact, only two 
changes are required.


(1) For the first term on the right hand side of  Eq. \ref{eq:dEdR_final}, one must  replace $\grad{R}E_k$ with
\begin{align}
    \grad{R}D^{kl} = \bm{0}.
\end{align}
The vanishing of the matrix element above
arises from the argument that $\bbm{C}$ is the independent variable and $\bbm{C}^0$ is a constant matrix. Therefore, we can rewrite Eq. \ref{eq:dEdR_final} explicitly for the purpose of computing derivative couplings as:
\begin{align}
    \dd{D^{kl}}{\bm{R}} = - (\grad{R}\bm{G})^\top\bm{\mu}-(\grad{R}\grad{A}E_{\rm tot})^\top\bm{\nu}+(\grad{R}\grad{A}G_0)^\top\bm{\nu}.\label{eq:dDkldR}
\end{align}

(2) In Eq. \ref{eq:munu_grad_adiabat}, one must replace $\grad{A}E_k$ with $\grad{A}D^{kl}$ where
\begin{align}
    \left(\grad{A}D^{kl}\right)_{p<q} = \begin{cases}
        1,& p = a_k, q = a_l, k<l\\
        -1,&p = a_l, q = a_k, k>l\\
        0,&{\rm otherwise}.
    \end{cases}
\end{align}
After solving for the Lagrange multipliers $\bm{\mu},\bm{\nu}$ from the revised Eq. \ref{eq:munu_grad_adiabat}, $\dd{D^{kl}}{\bm{R}}$ is computed through Eq. \ref{eq:dDkldR}, and finally the derivative coupling $\bm{d}^{kl}$ is computed through Eq. \ref{eq:d_kl_vfinal}.

\section{Numerical Results}\label{sec:result}

\begin{figure*}[ht]
    \centering\includegraphics[width=0.9\textwidth]{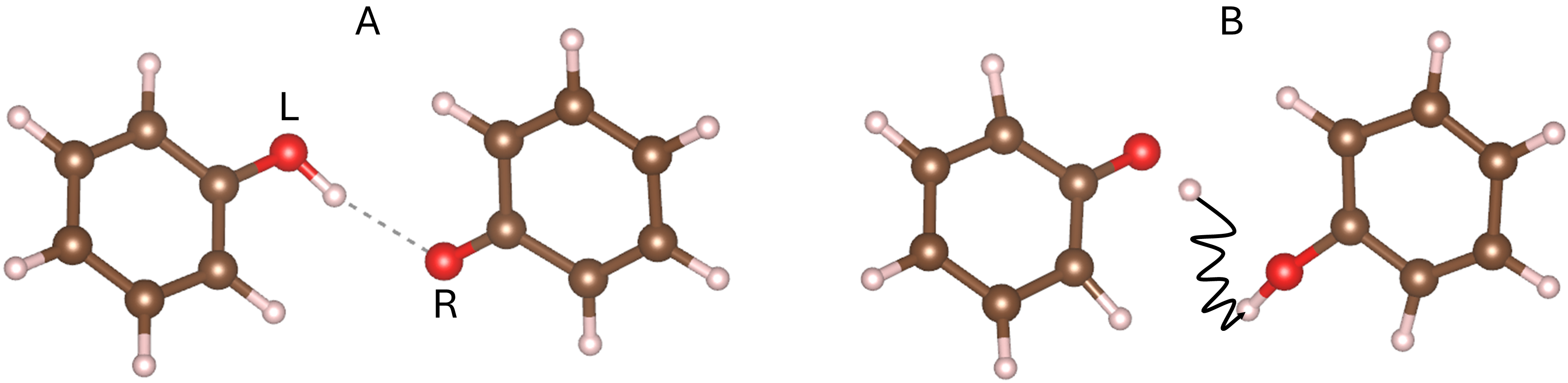}
    \caption{Geometry of the phenoxy-phenol system. (A) Initial geometry as relaxed to the minimum of the ground state potential energy surface in hDSC. ``L'' and ``R'' labels the oxygen atom from the ring on the left and right, respectively. (B) A schematic plot of the averaged hydrogen trajectory in SH simulations. The bridging hydrogen atom travels from the ring on the left to the ring on the right.}
    \label{fig:geometry}
\end{figure*}

The derivations in Secs. \ref{sec:ana_grad_ek} and \ref{sec:ana_dij} have been verified by comparing our computations with finite different results. As a proof-of-concept, we will now demonstrate the practical utility of our approach by using analytic nuclear gradients and derivative couplings to study a  proton-coupled electron transfer (PCET) process in a phenoxy-phenol system through a surface hopping (SH) simulation. \cite{tully:fssh} Our target is the  phenoxy-phenol molecule, which as been well studied by the Hammes-Schiffer group\cite{sirjoosingh:2011}.  To begin our analysis, using the analytic hDSC gradient,  we optimized the nuclear geometry of the complex and relaxed the system to its ground-state minimum.\cite{Qiu:2024:dsc} The geometry is shown in Fig. \ref{fig:geometry}A. 
Next, we excited the system to the first electronic excited state and ran SH dynamics as initialized with atomic velocities from a Maxwell-Boltzmann distribution corresponding to 300 K. The dynamics employed 100 independent trajectories, each propagated with a classical time step of 0.242 fs for a total simulation time of 60.5 fs. A two-quantum time step \cite{Qiu:2023:sh} scheme was applied to interpolate the Hamiltonian and the wavefunction propagation. 

\begin{figure*}[ht]
    \centering\includegraphics[width=1.0\textwidth]{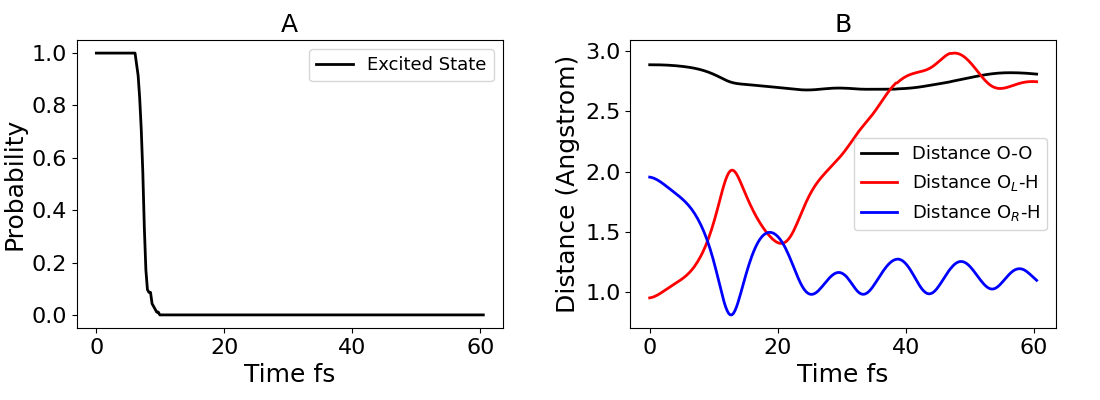}
    \caption{Dynamics of phenoxyl-phenol from SH simulations. (A) Excited state probability for the phenoxyl-phenol  complex as a function of time. All trajectories hop to ground state within 10 fs. (B) The ensemble averaged distances between two oxygen atoms (O-O), the oxygen atom from the left fragment and the bridging hydrogen atom (O$_{L}$-H), and the oxygen atom from the right fragment and the bridging hydrogen (O$_{R}$-H). The labeling of the oxygen atom is provided in Fig. \ref{fig:geometry}. The proton transfers from one ring to the other ring  within approximately 10 fs and never transfers back.}
    \label{fig:dynamics}
\end{figure*}

In Fig. \ref{fig:dynamics},  we plot the excited-state population $(A)$ and the ensemble average of three different bond lengths $(B)$. From our simulation, we find that proton transfer occurs in approximately 10 fs (see the red and blue curves in Fig. \ref{fig:dynamics}B). After the transfer, the bridging hydrogen atom never transfers back, but dissipates the energy into the phenoxy ring and oscillates around the oxygen atom. A schematic plot of the hydrogenic motion is shown in Fig. \ref{fig:geometry}B. Note that all trajectory hops from the excited state to the ground state occur in a short time duration that aligns with the proton transfer, in agreement with the notion that the proton transfers through a diabatic process.

As far as the efficiency of our implementation is concerned, each SCF calculation takes 25 seconds while each gradient calculation requires 34 seconds on a 48-core node. 14 seconds out of the 34 seconds needed for gradient calculations are spent on preparing basis function integrals and tensor contractions engine (such as an engine to calculate the contraction between $\bm{\Pi}^\RR$ and the density matrix), which are prime candidates for further optimization in the near future. Even with these inefficiencies, however, we find the complete surface hopping simulation for the phenoxy-phenol system requires approximately 1 minute per classical time step and 4 hours per trajectory.  Thus, the present simulations demonstrates that our method provides a computationally feasible approach for studying CT processes and nonadiabatic dynamics in systems of practical interest going forward.

\section{Discussion and Conclusions}\label{sec:conclusion}
In this paper, we have derived and implemented the analytic nuclear gradient and derivative coupling for the recently developed $(1,M)$ and $(2M-1,M)$ dynamically-weighted  constrained CASSCF method (so-called eDSC/hDSC)  through a  Lagrangian formalism. Our implementation provides an efficient way to study CT processes with globally smooth surfaces, making nonadiabatic dynamics simulations of large molecular systems feasible for short and long times. This work can be immediately applied to various research fields, including the study of excited-state dynamics in photocatalytic systems with explicit treatment of quantum effects at avoided crossings, modeling electron transfer reactions at electrode interfaces, describing the multi-reference character of transition metal complexes during catalytic cycles, simulating the charge transport mechanism in molecular junctions, as well as many other physical processes.

Future directions for this work include further optimization of the integral computation routines, which currently account for a significant portion of the gradient calculation time. We are also extending eDSC/hDSC to incorporate the generalized Hartree-Fock (GHF) formalism with spin-orbit coupling (SOC), enabling the study of spin dynamics in radical systems where both charge transfer and spin evolution play critical roles. There is today a great deal of research pointing out that electron transfer may well be spin-polarized -- even in organic systems without much spin polarization.\cite{naaman:2019:natrev}  To that end, a GHF framework for eDSC/hDSC calculations will serve as a zeroth order algorithm for implementing a phase-space Hamiltonian approach that couples nuclear motion with electronic and spin degrees of freedom and propagates electron transfer and nonadiabatic dynamics in a manner that  conserves the total linear and angular momentum\cite{Bian2024,Tao2025}. Thus, at the end of day, our hope is that the present algorithm should be an initial step in a exciting search for efficient and accurate large-scale nonadiabatic  simulations of electron transfer in complex molecular systems.

\section{Acknowledgments}

\section{Appendix}
\def\thesubsection{\thesection.\Alph{subsection}}

\setcounter{equation}{0}
\def\theequation{A\arabic{equation}}
\subsection[ddEtot]{Second Differential of $E_{\rm tot}$}\label{sec:Appendix:wi}
In this section we derive the second differential of $E_{\rm tot}$ given the dynamical weighting functions in Eqs. \ref{eq:w_first}-\ref{eq:w_last}. We start with $a_{j>1} = \frac{1-e^{-\beta \Delta E_j}}{\beta \Delta E_j}$ (note that $a_1=1$), which gives
\begin{align}
    d(a_j\Delta E_j) &= e^{-\beta \Delta E_j}d\Delta E_j,\\
    \Rightarrow da_j &= \frac{e^{-\beta \Delta E_j}-a_j}{\Delta E_j}d\Delta E_j.
\end{align}
We can define
\begin{align}
     a_{j>1}' = \frac{e^{-\beta \Delta E_j}-a_j}{\Delta E_j},
\end{align}
such that
\begin{align}
    da_j = a_j'd\Delta E_j.
\end{align}
Next,
\begin{align}
    d(a_j'\Delta E_j) &= -\beta e^{-\beta\Delta E_j}d\Delta E_j-a_j'd\Delta E_j,\\
    \Rightarrow d(a_j')&=-\frac{\beta e^{-\beta\Delta E_j}+2a_j'}{\Delta E_j}d\Delta E_j \equiv a_j''d\Delta E_j,
\end{align}
where we have defined
\begin{align}
    a_{j>1}'' = -\frac{\beta e^{-\beta\Delta E_j}+2a_j'}{\Delta E_j}.
\end{align}
Now, instead of working with $E_{\rm tot}$, it is beneficial to define
\begin{align}
    \Delta E = E_{\rm tot}-E_1 = \frac{\sum_{i>1} a_i\Delta E_i}{\sum_ja_j},
\end{align}
such that one has 
\begin{align}
    d(\Delta E\sum_{j}a_j) &=\sum_{i>1}e^{-\beta\Delta E_i}d\Delta E_i,\\
    \Rightarrow d(\Delta E) &= \frac{\sum_{i>1} \left(e^{-\beta \Delta E_i}-a_i'\Delta E\right)d\Delta E_i}{\sum_j a_j}\equiv \sum_{i>1} b_i'd\Delta E_i,
\end{align}
where we have defined
\begin{align}
    b_{i>1}' = \frac{\left(e^{-\beta \Delta E_i}-a_i'\Delta E\right)}{\sum_j a_j}.
\end{align}
Finally,
\begin{align}
    d^2(\Delta E) = &\frac{1}{\sum_j a_j}\left(\sum_{i>1}\left(-\beta e^{-\beta\Delta E_i}d\Delta E_i -a_i''\Delta Ed\Delta E_i - a_i'd\Delta E\right)d\Delta E_i\right)\nonumber\\
    -&\frac{1}{\left(\sum_j a_j\right)^2}\left(\sum_{i>1} \left(e^{-\beta \Delta E_i}-a_i'\Delta E\right)d\Delta E_i\right)\left(\sum_{j>1}a_j'd\Delta E_j\right)+\sum_{i>1}b_i'd^2\Delta E_i\\
    =&\frac{\sum_{i>1}\left(-\beta e^{-\beta\Delta E_i}-a_i''\Delta E\right)(d\Delta E_i)^2}{\sum_j a_j}-\frac{\sum_{ij>1}(a_i'b_j'+a_j'b_i')d\Delta E_id\Delta E_j}{\sum_j a_j}+\sum_{i>1}b_i'd^2\Delta E_i.
\end{align}
If we define
\begin{align}
    z_{i\neq j} &= -\frac{a_i'b_j'+a_j'b_i'}{\sum_ja_j},\\
    z_{ii} &=\frac{-\beta e^{\beta \Delta E_i}-a_i''\Delta E-2a_i'b_i'}{\sum_ja_j}.
\end{align}
then we can write
\begin{align}
    d^2(E_{\rm tot}) = \sum_{ij>1}z_{ij}d\Delta E_id\Delta E_j+\sum_{j>1}b_j'd^2\Delta E_j + d^2E_1.
\end{align}

\setcounter{equation}{0}
\def\theequation{B\arabic{equation}}
\subsection[TR]{Nuclear Gradient of $\bm{T}=\bm{S}^{1/2}$}\label{sec:Appendix:TR}
To compute $\bm{T}^\RR$ from $\bm{S}^\RR$, note that
\begin{align}
    \bm{T}^\RR\bm{T} + \bm{T}\bm{T}^\RR = \bm{S}^\RR.
\end{align}
Now, let us suppose we diagonalize $\bm{T}$:
\begin{align}
    \bm{T} = \bm{U\Lambda U}^\top,
\end{align}
Therefore, using the above two equations, it follows that:
\begin{align}
    \left(\bm{U}^\top\bm{T}^\RR\bm{U}\right)\bm{\Lambda} + \bm{\Lambda}\left(\bm{U}^\top\bm{T}^\RR\bm{U}\right) &= \bm{U}^\top\bm{S}^\RR\bm{U}\\
    \Rightarrow \bm{V\Lambda}+\bm{\Lambda V}&= \bm{U}^\top\bm{S}^\RR\bm{U},
\end{align}
where
\begin{align}
    \bm{V} = \bm{U}^\top\bm{T}^\RR\bm{U}.
\end{align}
Since $\bm{\Lambda}$ is diagonal, the solution to $\bm{V}$ is straightforward,
\begin{align}
    V_{ij} = \left(\bm{U}^\top\bm{S}^\RR\bm{U}\right)_{ij} / \left(\Lambda_{ii}+\Lambda_{jj}\right),
\end{align}
and finally, 
\begin{align}
    \bm{T}^\RR = \bm{UVU}^\top.
\end{align}

\bibliography{cite.bib}
\end{document}